\documentclass[structabstract]{aa}  

\usepackage{graphicx}
\usepackage{pstricks}

\usepackage{txfonts}
\usepackage{natbib}
  \bibpunct{(}{)}{,}{a}{}{,}

\usepackage{hyperref}
\hypersetup{
    bookmarks=true,         
    unicode=false,          
    pdftoolbar=true,        
    pdfmenubar=true,        
    pdffitwindow=false,     
    pdftitle={The Family of (136108) Haumea},
    pdfauthor={Carry et al.}, 
    pdfsubject={Planetary Science}, 
    pdfkeywords={},         
    pdfnewwindow=true,      
    colorlinks=true,        
    linkcolor=gray,         
    citecolor=blue,         
    filecolor=gray,         
    urlcolor=gray           
}

\usepackage{ulem}
\newcommand{\add}[1]{\textcolor{blue}{\bf{#1}}}

\renewcommand{\add}[1]{#1}

\newcommand{\ie}{\textsl{i.e.}}
\newcommand{\eg}{\textsl{e.g.}}

\begin{document}

  \title{Characterisation of candidate members of (136108) Haumea's family
    \thanks{Based on observations
      collected at the European Southern Observatory, La Silla \& Paranal, Chile -
      \href{http://archive.eso.org/wdb/wdb/eso/sched_rep_arc/query?progid=81.C-0544}%
           {81.C-0544}
    \& 
      \href{http://archive.eso.org/wdb/wdb/eso/sched_rep_arc/query?progid=82.C-0306}%
           {82.C-0306}
    \& 
      \href{http://archive.eso.org/wdb/wdb/eso/sched_rep_arc/query?progid=84.C-0594}%
           {84.C-0594}
  }}

  \subtitle{II. Follow-up observations}
  
  \author{Beno\^{i}t Carry\inst{1,2,3,4}
    \and Colin Snodgrass\inst{5,6}
    \and Pedro Lacerda\inst{7}
    \and Olivier Hainaut\inst{8}
    \and Christophe Dumas\inst{6}
  }
  
  \offprints{B. Carry, \email{benoit.carry@esa.int}}
  
  \institute{
    European Space Astronomy Centre, ESA,
    P.O. Box 78, 28691 Villanueva de la Ca\~{n}ada, Madrid, Spain
    \and
    IMCCE, Observatoire de Paris, UPMC, CNRS, 77 Av. Denfert Rochereau 75014 Paris, France
    \and
    LESIA, Observatoire de Paris-Meudon,
    5 place Jules Janssen, 92195 Meudon Cedex, France
    \and
    Universit\'e Paris 7 Denis-Diderot,
    5 rue Thomas Mann,
    75205 Paris CEDEX, France
    \and
    Max Planck Institute for Solar System Research, Max-Planck-Strasse 2,
    37191 Katlenburg-Lindau, Germany
    \and
    European Southern Observatory, Alonso de C\'{o}rdova 3107,
    Vitacura, Casilla 19001, Santiago de Chile, Chile
    \and
    Queen's University, Belfast, County Antrim BT7 1NN, Northern Ireland
    \and
    European Southern Observatory,
    Karl-Schwarzschild-Strasse 2,
    D-85748 Garching bei M\"{u}nchen, Germany
  }

\date{Received  / Accepted }

\abstract
{From a dynamical analysis of the orbital elements of transneptunian
  objects (TNOs), Ragozzine \& Brown reported a list of
  candidate members of the first collisional family found
  among this population, associated with (136\,108) Haumea
  (a.k.a. 2003~EL$_{61}$).} 
{We aim to distinguish the true members of the Haumea
  collisional family from interlopers. We search for water ice on
  their surfaces, which is a common characteristic of the
  known family members. The properties of the confirmed family are
  used to 
  constrain the formation mechanism of Haumea, its satellites, and its family.} 
{Optical and near-infrared photometry is used to identify
  water ice. We use in particular the $CH_4$ filter of
  the Hawk-I instrument at the European Southern Observatory
  Very Large Telescope as a short $H$-band ($H_S$),
  the $(J - H_S)$ colour being a sensitive measure of
  the water ice absorption band at 1.6 $\mu m$.} 
{Continuing our previous study headed by Snodgrass, we 
  report colours for 8 candidate family members, including 
  near-infrared colours for 5.
  We confirm one object as a genuine member of the collisional
  family (2003~UZ$_{117}$), and reject 5 others.
  The lack of infrared data for the two remaining objects prevent any
  conclusion from being drawn.
  The total number of rejected members is therefore 17. The 11 
  confirmed members represent only a third of the 36 candidates.}
{The origin of Haumea's family is likely to be related to an
  impact event. However, a scenario explaining all the peculiarities
  of Haumea itself and its family remains elusive.}


\keywords{Kuiper Belt; 
Methods: observational;
Techniques: photometric;
Infrared: solar system}

   \maketitle
%


\section{Introduction}

  \indent The dwarf planet (136\,108) Haumea
  \citep{2005-IAUC-8577-Santos-Sanz} is among the largest
  objects found in the Kuiper Belt 
  \citep{2006-ApJ-639-Rabinowitz, 2008-SSBN-3-Stansberry},
  together with Pluto, Eris, and Makemake.
  It is a highly unusual body with the following characteristics:
  \begin{enumerate}
    \item It has a very elongated cigar-like shape
      \citep{2006-ApJ-639-Rabinowitz, 2010-AA-518-Lellouch}.

    \item It is a fast rotator \citep[$P_{rot} \sim 3.9$\,h,][]{2006-ApJ-639-Rabinowitz}.

    \item It has two non-coplanar satellites
      \citep{2006-ApJ-639-Brown, 2009-AJ-137-Ragozzine, 2011-AA-528-Dumas}.

    \item It is the largest member of a dynamical family
      \citep{2007-Nature-446-Brown, 2007-AJ-134-Ragozzine}, whose
      velocity dispersion is surprisingly small
      \citep{2009-ApJ-700-Schlichting, 2010-ApJ-714-Leinhardt}.

    \item Its surface composition is dominated by water ice
      \citep{2007-AJ-133-Tegler, 2007-ApJ-655-Trujillo,
        2007-AA-466-Merlin, 2009-AA-496-Pinilla-Alonso, 2011-AA-528-Dumas},
      yet it has a high density of 2.5-3.3\,g\,cm$^{-3}$
      \citep{2006-ApJ-639-Rabinowitz}.

    \item It surface has a hemispherical colour heterogeneity, 
      with a dark red ``spot'' on one side
      \citep{2008-AJ-135-Lacerda, 2009-AJ-137-Lacerda}.

  \end{enumerate}

  \indent \citet{2007-Nature-446-Brown} 
  proposed that Haumea suffered a giant collision
  that ejected a large fraction of its ice mantle,
  which formed both the two satellites and the
  dynamical family
  and left Haumea with rapid rotation.
  A number of theoretical studies have since looked at the family 
  formation in more detail (see Sect.~\ref{sec: discussion}). \\
  \indent A characterisation of the candidate members
  \citep[35 bodies listed by][including Haumea
    itself]{2007-AJ-134-Ragozzine} 
  however showed that only 10 bodies out of 24 studied share their
  surface properties with Haumea \citep{2010-AA-511-Snodgrass}, and
  can thus be considered genuine family members.
  Moreover, these confirmed family members cluster in the
  orbital elements space
  \citep[see Fig.~4 in][]{2010-AA-511-Snodgrass}, and the 
  highest velocity found was $\sim$123\,m\,s$^{-1}$ (for 1995 SM$_{55}$).\\
  \indent We report on follow-up observations to
  \citet{2010-AA-511-Snodgrass}
  of 8 additional candidate members of Haumea's family.
  We describe our observations in Sect.~\ref{sec: obs}, 
  the colour measurements in Sect.~\ref{sec: colours}, 
  the lightcurve analysis and density estimates in 
  Sect.~\ref{sec: dens}, and
  we discuss in Sect.~\ref{sec: discussion} the family memberships of
  the candidates and the implication of these for the characteristics
  of the family.

\section{Observations and data reduction\label{sec: obs}}
  \indent We performed our observations
  at the European Southern
  Observatory (ESO)
  La Silla and Paranal
  Very Large Telescope (VLT) sites (programme ID:
  \href{http://archive.eso.org/wdb/wdb/eso/sched_rep_arc/query?progid=84.C-0594}%
       {84.C-0594}).
  Observations in the visible wavelengths ($BVRi$ filters) were performed 
  using the EFOSC2 instrument
  \citep{1984-Msngr-38-Buzzoni} mounted on the NTT
  \citep[since April 2008;][]{2008-Msngr-132-Snodgrass};
  while near-infrared observations ($J$, $CH_4$ filters)
  were performed using the wide-field camera Hawk-I
  \citep{2004-SPIE-5492-Pirard, 2006-SPIE-6269-Casali,
    2008-AA-491-Kissler-Patig} installed on 
  the UT4/Yepun telescope.
  We use the medium-width $CH_4$ filter as a
  narrow H band
  (1.52--1.63\,$\mu$m, hereafter $H_S$)
  to measure the $J$-$H_S$ colour as
  a sensitive test for water ice
  \citep[see][for details]{2010-AA-511-Snodgrass}.
  We list the observational circumstances in Table~\ref{tab: obs}.\\ 
  \indent We reduced the data in the usual manner
  (\ie, bias subtraction, flat fielding,
  sky subtraction,~as appropriate). 
  We refer readers to 
  \citet{2010-AA-511-Snodgrass} for a complete description of
  the instruments and the methods we used
  to detect the targets, and both measure and calibrate their
  photometry. \\ 
  \indent For each frame, we used the SkyBoT cone-search
  method \citep{2006-ASPC-351-Berthier}
  to retrieve all known solar system objects
  located in the field of view.
  We found 3 main-belt asteroids,
  and the potentialy hazardeous asteroid (29\,075) 1950 DA
  \citep[\textsl{e.g.}][]{2002-Science-296-Giorgini,
    2003-GeoJI-153-Ward}, in our frames.
  We report
  the circumstances of their serendipitous observations in
  Table~\ref{tab: obs} and their apparent magnitude in
  Table~\ref{tab: phot}, together with the family candidates and our
  back-up targets.

\begin{table}[h!]
  \caption{Observational circumstances.}
  \label{tab: obs}
\centering
\begin{tabular}{r@{ ~}lrrrcl}
  \hline\hline
  \multicolumn{2}{c}{Object} &
  \multicolumn{1}{c}{$\Delta^{\mathrm{a}}$}  &
  \multicolumn{1}{c}{$r^{\mathrm{b}}$} &
  \multicolumn{1}{c}{$\alpha^{\mathrm{c}}$} &
  \multicolumn{1}{c}{Runs$^{\mathrm{d}}$} \\
  \multicolumn{1}{c}{(\#)} &
  \multicolumn{1}{c}{(Designation)} &
  \multicolumn{1}{c}{(AU)} &
  \multicolumn{1}{c}{(AU)} &
  \multicolumn{1}{c}{(\degr)} & \\
  \hline
  \noalign{\smallskip}
         & 1999 CD 158 & 47.5 & 46.5 &  0.5 & B \\
         & 1999 OK 4   & 46.5 & 45.5 &  0.3 & $\star$ \\
         & 2000 CG 105 & 45.8 & 46.8 &  0.1 & A,B \\
         & 2001 FU 172 & 32.2 & 32.0 &  1.7 & A \\
         & 2002 GH 32  & 43.2 & 42.9 &  1.2 & B \\
         & 2003 HA 57  & 32.7 & 32.3 &  1.6 & A \\
         & 2003 UZ 117 & 39.4 & 39.4 &  1.4 & A \\
         & 2004 FU 142 & 33.5 & 33.2 &  0.0 & A \\
         & 2005 CB 79  & 39.9 & 39.0 &  0.4 & A \\
         & 2005 GE 187 & 30.3 & 30.2 &  1.9 & A \\
  \noalign{\smallskip}
 \hline
 \noalign{\smallskip}
      24 & Themis      &  3.4 &  4.0 & 12.0 & B \\
 10\,199 & Chariklo    & 13.8 & 13.6 &  4.1 & B \\
 29\,075 & 1950 DA     &  0.8 &  1.0 & 62.7 & A \\
158\,589 & Snodgrass   &  3.5 &  3.1 & 15.5 & A \\
104\,227 & 2000 EH 125 &  3.0 &  2.5 & 18.5 & A \\ 
202\,095 & 2004 TQ 20  &  2.2 &  1.9 &  2.4 & A \\
         & 2010 CU 19  &  1.3 &  1.6 &  0.6 & A \\
  \noalign{\smallskip}
  \hline
\end{tabular}
\tablefoot{
$^{\mathrm{(a)}}$ Heliocentric distance.
$^{\mathrm{(b)}}$ Geocentric distance.
$^{\mathrm{(c)}}$ Phase angle.
$^{\mathrm{(d)}}$ Runs: A = 2010 February 15--17, EFOSC2;
  B= 2010 February 22, Hawk-I.
$^{\star}$ Observed on 2009 July 24 with EFOSC2.
}
\end{table}


\section{Colours\label{sec: colours}}
\begin{figure*}
   \centering
   \includegraphics[angle=270, width=\textwidth]{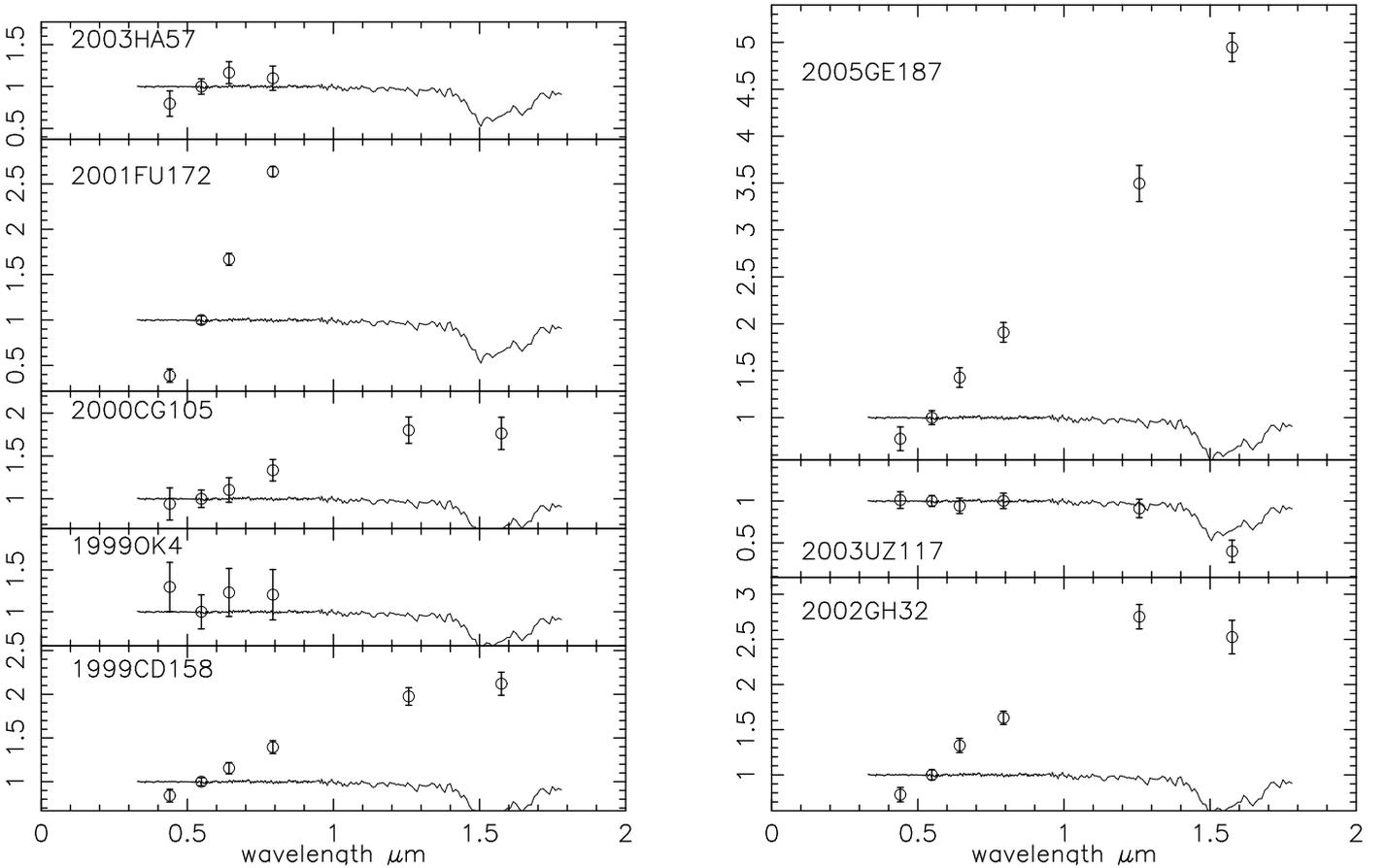} 
   \caption[VNIR spectra of family candidates]{%
     Visible and
     near-infrared photometry for the candidate family 
     members (see Table~\ref{tab: colors}).
     The data are normalized at 0.55\,$\mu$m ($V$ filter).
     The spectrum of Haumea
     \citep[taken from][]{2009-AA-496-Pinilla-Alonso}
     is shown for comparison in each.}
   \label{fig: spec}
\end{figure*}

  \indent We report the photometry of all the objects
  in Table~\ref{tab: phot}, where we give
  the apparent magnitude in each band, averaged over all the observations.
  We used a common sequence of filters (RBViR) to observe all the
  objects.
  This limits the influence of the shape-related lightcurve on the
  colour determination.
  In Table~\ref{tab: colors}, we report the average colours of all the family
  candidates observed here, and refer to
  \citet{2010-AA-511-Snodgrass} for a complete review of the
  published photometry.\\
%
\begin{table*}
\caption{Mean apparent magnitudes for each object.}
  \label{tab: phot}
\centering
\begin{tabular}{l cccc cc}
  \hline\hline
  \multicolumn{1}{c}{Object} & $B$  & $V$ & $R$ & $i$ & $J$ & $CH_4$ \\
  \hline
  \noalign{\smallskip}
  1999 CD 158 &         --       &         --       &         --       &         --       & 20.79 $\pm$ 0.08 & 20.44 $\pm$ 0.10  \\
  1999 OK 4      & 24.90 $\pm$ 0.16 & 24.54 $\pm$ 0.17 & 23.95 $\pm$ 0.14 & 23.64 $\pm$ 0.20 &     --    &     --     \\
  2000 CG 105 & 24.32 $\pm$ 0.14 & 23.62 $\pm$ 0.10 & 23.15 $\pm$ 0.05 & 22.61 $\pm$ 0.07 & 21.89 $\pm$ 0.10 & 21.64 $\pm$ 0.14  \\
  2001 FU 172 & 23.40 $\pm$ 0.05 & 21.73 $\pm$ 0.04 & 20.82 $\pm$ 0.03 & 19.99 $\pm$ 0.03 &         --       &         --        \\
  2002 GH 32  &         --       &         --       &         --       &         --       & 21.49 $\pm$ 0.12 & 21.31 $\pm$ 0.15 \\
  2003 HA 57  & 24.37 $\pm$ 0.09 & 23.48 $\pm$ 0.09 & 22.96 $\pm$ 0.05 & 22.69 $\pm$ 0.12 &         --       &         --      \\
  2003 UZ 117$^\dagger$ & 21.86 $\pm$ 0.09 & 21.34 $\pm$ 0.08 & 21.09 $\pm$ 0.08 & 20.67 $\pm$ 0.07 & -- & --\\ 
  2003 UZ 117$^\star$ & 22.04 $\pm$ 0.10 & 21.32 $\pm$ 0.06 & 21.01 $\pm$ 0.06 & 20.62 $\pm$ 0.06   & -- & -- \\ 
  2005 CB 79  &         --       &         --       & 20.29 $\pm$ 0.01 &         --       &         --       &         --        \\
  2005 GE 187 & 23.73 $\pm$ 0.14 & 22.91 $\pm$ 0.12 & 22.23 $\pm$ 0.09 & 21.49 $\pm$ 0.06 &         --       &         --        \\
  \noalign{\smallskip}
  \hline
  \noalign{\smallskip}
  1950 DA     & 19.59 $\pm$ 0.07 & 19.15 $\pm$ 0.06 & 18.82 $\pm$ 0.02 & 18.56 $\pm$ 0.04 &         --       &         --       \\
  2000 EH 125 & 21.58 $\pm$ 0.03 & 20.78 $\pm$ 0.02 & 20.37 $\pm$ 0.02 & 20.05 $\pm$ 0.03 &         --       &         --       \\ 
  2004 TQ 20  & 21.93 $\pm$ 0.06 & 21.23 $\pm$ 0.07 & 21.19 $\pm$ 0.08 & 20.73 $\pm$ 0.07 &         --       &         --       \\
  2010 CU 19  &         --       & 19.26 $\pm$ 0.04 &         --       &         --       &         --       &         --       \\
  Chariklo    &         --       &         --       &         --       &         --       & 16.98 $\pm$ 0.02 & 16.86 $\pm$ 0.02 \\
  Themis      &         --       &         --       &         --       &         --       & 12.38 $\pm$ 0.02 & 12.25 $\pm$ 0.02 \\
  Snodgrass   & 22.40 $\pm$ 0.14 & 21.61 $\pm$ 0.10 & 21.20 $\pm$ 0.05 & 20.69 $\pm$ 0.08 &         --       &         --       \\
  \noalign{\smallskip}
  \hline
\end{tabular}
\tablefoot{$^\dagger$ First night, $^\star$ Second night.}
\end{table*}
%
%
%
  \indent From these average colours, we calculate reflectances by
  comparing them to the solar colours. 
  We also report the visible slope for each object
  (\%/100 nm) in Table~\ref{tab: colors},
  calculated from the reflectances via a linear regression over the
  full $BVRi$ range.
  The reflectance ``spectra'' of the candidates from
  this photometry are shown in Fig.~\ref{fig: spec}.
  The reflectance spectrum of (136\,108) Haumea from
  \citet{2009-AA-496-Pinilla-Alonso} is shown for comparison to the
  photometry.
  For all the objects but 
  1999 CD$_{158}$ \citep{2004-AA-417-Delsanti},
  the link between
  the visible and near-infrared wavelengths was made by extrapolating
  the visible spectral slope to the $J$-band, owing to a lack of simultaneous
  observations. 
  Among these objects, 2002 GH$_{32}$ has a distinctive spectral
  behaviour. It displays a slight dip at 1.5\,$\mu$m despite a
  red slope, as its $(J-H_S)$ colour (0.18\,$\pm$\,0.19) is slightly bluer
  than that of the Sun \citep[0.28;][]{2010-AA-511-Snodgrass}.
  Given the uncertainty in
  this point, and the red optical slope, we do not believe that this is
  evidence of strong water ice absorption. \\
  \indent From the visible and near-infrared colours that we report here,
  we confirm that 2003~UZ$_{117}$ is a genuine family member, 
  in agreement with \citet{2007-AJ-134-Ragozzine} and \citet{2010-AA-511-Snodgrass},
  and reject 
  1999~CD$_{158}$, 
  2000~CG$_{105}$, 
  2001~FU$_{172}$,
  2002~GH$_{32}$, and
  2005~GE$_{187}$.
  The TNO 1999 OK$_{4}$ remains a possible candidate, as it has a neutral slope
  in the visible, but the poor
  signal-to-noise ratio of the data for this faint target does not
  allow us to draw a stronger conclusion.
  In any case, a neutral slope by itself does not
  confirm family membership without near-infrared observations.
  This object is dynamically very near to the centre of the family and
  remains worthy of further investigation.
  2003 HA$_{57}$ has a red slope, but not a
  very strong one. It is further from the centre of the distribution,
  with $\delta v > 200$\,m\,s$^{-1}$, so it is unlikely to be a family
  member (see below).
  We cannot firmly conclude anything about the membership of 1999 OK$_{4}$
  and 2003 HA$_{57}$.
  The current number of confirmed family members is
  11 over 36 (including Haumea and an
  additional dynamical candidate (2009 YE$_{7}$) that was found and directly
  confirmed by \citet{2011-ApJ-730-Trujillo}), or 31\%. 
  The number of rejected candidates is 17 over 36, hence 47\% of the
  population, and there are only 8 objects whose status remains
  unknown. 

%
\begin{table*}
\caption[Average colours of the family candidates]{Average colours in $BVRiJH_S$, and assessment of likely membership based on these colours.}
  \label{tab: colors}
\centering
\begin{tabular}{lcccccccc}
  \hline\hline
  \multicolumn{1}{c}{Object} & $(B-V)$ & $(V-R)$ & $(R-i)$ & $(R-J$) & $(J-H_S)^\star$ & Vis.~slope &Ref. & Family?\\
  \multicolumn{1}{c}{Designation} & (mag.) & (mag.) & (mag.) & (mag.) & (mag.) & (\%/100nm) & & \\
  \hline
  \noalign{\smallskip}
  1999 CD 158 & 0.83 $\pm$ 0.06 & 0.51 $\pm$ 0.05 & 0.54 $\pm$ 0.06 & 1.38 $\pm$ 0.09 & 0.35 $\pm$ 0.12 & 15.8 $\pm$ 0.6  & 1,5,8 &N\\
  1999 OK 4   & 0.36 $\pm$ 0.23 & 0.58 $\pm$ 0.22 & 0.32 $\pm$ 0.24 &        --       &        --       &  1.4 $\pm$ 18.1 &  8  & ? \\
  2000 CG 105 & 0.71 $\pm$ 0.17 & 0.56 $\pm$ 0.11 & 0.77 $\pm$ 0.29 &        --       & 0.25 $\pm$ 0.17 & 11.3 $\pm$  4.3 & 5,8 & N\\
  2000 JG 81  &        --       & 0.50 $\pm$ 0.11 & 0.33 $\pm$ 0.12 &        --       &        --       &  5.6 $\pm$ 21.6 & 6 & ? \\
  2001 FU 172 & 1.67 $\pm$ 0.06 & 0.91 $\pm$ 0.05 & 0.83 $\pm$ 0.03 &        --       &        --       & 64.2 $\pm$  4.3 & 5,8 & N\\
  2002 GH 32  & 0.91 $\pm$ 0.06 & 0.66 $\pm$ 0.06 &  0.56 $\pm$ 0.05&        --       & 0.18 $\pm$ 0.19 & 24.8 $\pm$  4.7 & 5,8 & N\\
  2003 HA 57  & 0.89 $\pm$ 0.13 & 0.52 $\pm$ 0.10 & 0.27 $\pm$ 0.12 &        --       &        --       &  8.7 $\pm$ 11.6 & 8 & ?\\
  2003 UZ 117 & 0.52 $\pm$ 0.12 & 0.25 $\pm$ 0.11 & 0.42 $\pm$ 0.11 &        --       &-0.74 $\pm$ 0.16 & -0.5 $\pm$  3.7 & 2-5,7,8 & Y \\
  2005 GE 187 & 0.81 $\pm$ 0.18 & 0.69 $\pm$ 0.14 & 0.74 $\pm$ 0.11 & 1.22 $\pm$ 0.19 & 0.65 $\pm$ 0.14 & 32.8 $\pm$ 12.3 & 5,8 & N \\
  \noalign{\smallskip}
  \hline
  \noalign{\smallskip}
  Haumea         & 0.64 $\pm$ 0.01 & 0.33 $\pm$ 0.01 & 0.34 $\pm$ 0.01 & 0.88 $\pm$ 0.01 & -0.60 $\pm$ 0.11 & -0.6  $\pm$ 0.9  & 5 & Y \\
  \noalign{\smallskip}
  \hline
\end{tabular}
\tablebib{%
[1] \citet{2004-AA-417-Delsanti};
[2] \citet{2009-AA-493-DeMeo};
[3] \citet{2007-AA-468-Pinilla-Alonso};
[4] \citet{2008-AA-487-Alvarez-Candal};
[5] \citet[][and references therein]{2010-AA-511-Snodgrass};
[6] \citet{2011-Icarus-213-Benecchi};
[7] \citet{2011-ApJ-730-Trujillo};
[8] This work.
Where colours for a given object are published by multiple authors, we
quote a weighted mean.
}
\tablefoot{$^\star$ In the present study, $H_s$ correspond to Hawk-I $CH_4$
filter}
\end{table*}

\section{Rotation and density\label{sec: dens}}

\begin{figure}
   \centering
   \includegraphics[width=.49\textwidth]{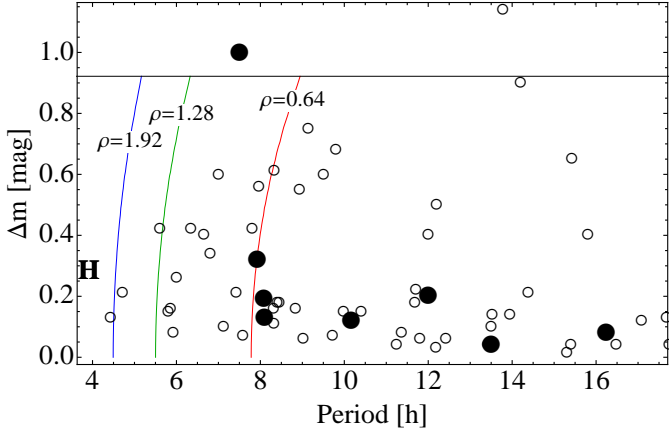} 
   \caption[Lightcurve amplitude vs. period]{%
     Lightcurve amplitude ($\Delta m$) as a function of the
     rotation period (in hours) for the TNOs in the vicinity of
     Haumea.
     Filled and open circles stand for confirmed family members and
     background population
     \citep[from][]{2009-AA-505-Duffard, 2010-AA-522-Thirouin},
     respectively. The letter H shows the position of Haumea.
     Vertical blue, red, and green curves are the limit for stability,
     assuming the objects are in hydrostatic equilibrium, \ie,
     stable objects left of a line are denser than the number in the
     label (in g\,cm$^{-3}$).
     Objects above the black line ($\Delta m$\,$\sim$\,0.9\,mag)
     are unstable (under the hydrostatic equilibrium assumption), and
     are likely contact binaries. 
   }
   \label{fig: dens}
\end{figure}

  \indent To constrain the density of family members, and
  therefore test the hypothesis that they are formed of almost pure
  water ice, we investigated their rotational lightcurves.
  In the February 2010 observing run, we performed a time series of
  $R$-band photometry on 2005 CB$_{79}$,
  which was demonstrated to be a family
  member by \citet{2008-ApJ-684-Schaller} and
  \citet{2010-AA-511-Snodgrass}.
  We measured 69 points over the course of three nights, with a
  typical uncertainty in each measurement of 0.03 magnitudes. We
  observed a variation of around 0.15 magnitudes, but found no
  convincing periodicity. \citet{2010-AA-522-Thirouin} found a period
  of 6.76 hours and a similar magnitude range. \\
  \indent A total of 8 family members have published lightcurve
  measurements  
  (Table~\ref{tab: period}). These can be used to
  estimate the density by two methods. By either balancing
  gravitational and centrifugal forces for an assumed strengthless
  (rubble pile) body, as applied to asteroids
  \citep{2002-AsteroidsIII-2.2-Pravec} and comets
  \citep{2006-MNRAS-373-Snodgrass}, or by assuming a fluid equilibrium
  shape (\ie, a Jacobi ellipsoid), which may be more appropriate for large
  icy bodies such as TNOs \citep{2007-AJ-133-Lacerda}. The 
  densities of TNOs derived from lightcurves was reviewed by
  \citet{2009-AA-505-Duffard} and \citet{2010-AA-522-Thirouin}. Of
  particular interest is the high value of 2.38 g\,cm$^{-3}$
  determined for 2003 OP$_{32}$,
  which is a large confirmed family member with a strong water-ice
  spectrum \citep{2007-Nature-446-Brown}. The quoted value
  is considerably higher than that of water ice, and close
  to the value determined for Haumea itself
  \citep[2.61\,g\,cm$^{-3}$,][]{2010-AA-522-Thirouin}, and is
  therefore inconsistent with this body being a pure water-ice
  fragment from the original Haumea's outer mantle. However, this
  (minimum) density is derived assuming that the best-fit single
  peaked period of 4.05 hours is the correct spin rate, which can only
  be true if the variation is due to an albedo patch on a spheroidal
  body, \ie, a Maclaurin spheroid rather than a Jacobi ellipsoid. If
  the true rotation period is instead twice this value (\ie, the
  double peaked lightcurve is due to shape instead of albedo
  features), then the
  required minimum density is 0.59\,g\,cm$^{-3}$, which provides a far
  weaker constraint. No other family member (aside from Haumea
  itself) has a reported rotation rate fast enough to require a high
  density (Table~\ref{tab: period} and Fig.~\ref{fig: dens}). \\
%
\begin{table*}
\caption{Rotational periods (SP: single peak, DP: double peak) of family candidates.}
  \label{tab: period}
\begin{center}
\begin{tabular}{rlcr c c@{\,$\pm$\,}c c@{\,$\pm$\,}ccc}
  \hline\hline
  \multicolumn{2}{c}{Object} & $H$ & $d^\dagger$ & $\Delta m$ &
     \multicolumn{2}{c}{Period SP} & \multicolumn{2}{c}{Period DP} &
     Ref. & $\rho_{m}^\star$ \\
  \# & Designation & & (km) & &\multicolumn{2}{c}{(h)} &
  \multicolumn{2}{c}{(h)} & & (g cm$^{-3}$)\\
  \hline
  \noalign{\smallskip}
 24835 & 1995 SM 55   & 4.8  & 174  & 0.19 &  4.04 & 0.03   &   8.08 & 0.03  & 2 & 0.60 \\
 19308 & 1996 TO 66   & 4.5  & 200  & 0.32 &  3.96 & 0.04   &   7.92 & 0.04  & 2 & 0.63 \\
       &              &      &      &      &  \multicolumn{2}{c}{ } & \multicolumn{2}{c}{11.9}  & 5 \\
       &              &      &      &      &  \multicolumn{2}{c}{ } &   6.25 & 0.03  & 1 \\
 86047 & 1999 OY 3    & 6.74 &  71  & \\
 55636 & 2002 TX 300  & 3.2  & 364  & 0.08 &  \multicolumn{2}{c}{8.16} & \multicolumn{2}{c}{  }  & 8 &0.16\\
       &              &      &      &      &  8.12 & 0.08   &  16.24 & 0.08  & 3 \\
       &              &      &      &      & 12.10 & 0.08   &  24.20 & 0.08  & 3 \\
       &              &      &      &      &  7.89 & 0.03   &  15.78 & 0.03  & 4 \\
136108 & Haumea       & 0.01 & 1313 & 0.28 &  \multicolumn{2}{c}{ } & 3.9154 & 0.0001  & 6,8,10 & 2.56\\
120178 & 2003 OP 32   & 3.95 & 258  & 0.13 &  \multicolumn{2}{c}{4.05} & \multicolumn{2}{c}{ }  & 8 & 0.59\\
       & 2003 SQ 317  & 6.3  &  87  & 1.00 &  3.74 & 0.10   &   7.48 & 0.10  & 9 & 0.5\\
       & 2003 UZ 117  & 5.3  & 138  & 0.20 &  \multicolumn{2}{c}{$\sim$6} &\multicolumn{2}{c}{ }  & 7 & 0.27\\
       & 2005 CB 79   & 4.7  & 182  & 0.04 &  \multicolumn{2}{c}{6.76} &\multicolumn{2}{c}{ }  & 8 & 0.21\\
145453 & 2005 RR 43   & 4.0  & 252  & 0.12 &  \multicolumn{2}{c}{7.87} & \multicolumn{2}{c}{ } & 8 & 0.38\\
       &              &      &      &      &  5.08 & 0.03  & \multicolumn{2}{c}{ } & 7 \\
       & 2009 YE 7    & 4.4  & 209  &\\
  \noalign{\smallskip}
  \hline
\end{tabular}
\tablebib{%
[1]~\citet{2000-AA-356-Hainaut};
[2]~\citet{2002-AJ-124-Sheppard};
[3]~\citet{2003-EMP-92-Sheppard};
[4]~\citet{2004-AA-420-Ortiz};
[5]~\citet{2006-Icarus-184-Belskaya};
[6]~\citet{2008-AJ-135-Lacerda};
[7]~\citet{2010-AA-510-Perna};
[8]~\citet{2010-AA-522-Thirouin};
[9]~\citet{2010-AA-511-Snodgrass};
[10]~\citet{2010-AA-518-Lellouch}}
\tablefoot{
  $^\dagger$ Diameter computed using an assumed geometric albedo of 0.7,
  with the 
  exception of Haumea, whose diameter is taken from
  \citet{2010-AA-518-Lellouch}. 
  2002 TX$_{300}$ has a diameter measurement of 286 km and albedo of 
  88\% \citep{2010-Nature-465-Elliot}, but these are inconsistent with
  the given $H$ magnitude. \\
  $^\star$ Density computed assuming a Jacobi ellipsoid shape
  with a DP rotation period (see text for details).}
\end{center}
\end{table*}
%
  \indent Instead of considering individual rotation periods, we
  consider the family as a whole. Fig.~\ref{fig: dens} compares all
  confirmed family members (black points) with all other TNO
  lightcurve measurements (open circles) taken from the compilation of 
  \citet{2009-AA-505-Duffard}. The rotation period plotted assumes a
  double-peaked period for all objects (\ie, shape-controlled
  lightcurve), and the curved lines show densities calculated based on
  the assumption of hydrostatic equilibrium (Jacobi ellipsoids). Rotation
  rates from the \citet{2009-AA-505-Duffard} table are taken at face
  value (no further attempt has been made to judge the reliability of
  the determined periods), with the exception of two very short
  rotation periods
  \citep[1996 TP$_{66}$ and 1998 XY$_{95}$, with single
  peak periods of 1.96 and 1.31 hours
  respectively;][]{1999-MNRAS-308-Collander-Brown,
    2001-MNRAS-325-Collander-Brown}  
  that appear in the table despite the original authors stating that
  these were unrealistic (and statistically insignificant)
  mathematical best fits. We removed these values and regard the
  rotation periods of these two objects as unknown. For all other
  objects where there are both multiple period determinations and no
  preferred period in \citeauthor{2009-AA-505-Duffard},
  we take the shortest period to give
  the highest possible minimum density.\\
%
%
  \indent Seven of the eight family members fall into the relatively
  long-period (low-density) area of this plot,
  with $\rho \le 0.64$\,g\,cm$^{-3}$.
  The exception is 2003 SQ$_{317}$, which has a large
  lightcurve amplitude \citep{2010-AA-511-Snodgrass}, implying that it
  is likely to be a contact binary
  \citep[therefore the Jacobi ellipsoid model does not
    hold,][]{2007-AJ-133-Lacerda}. \\
  \indent A direct comparison between the densities of family members
  and other 
  TNOs is not straightforward since analysis of the rotational
  properties based on hydrostatic equilibrium can in general only set
  lower limits on the densities of the objects. We can, however, use the
  observed lightcurve properties (Fig.~\ref{fig: dens}) to assess the
  probability that the family members and other TNOs were drawn from the
  same 2-D distribution in spin period vs. $\Delta m$.
  To do so, we use the 2-D Kolmogorov-Smirnov (K-S) test
  \citep{1983-MNRAS-202-Peacock}. The 2-D K-S test
  uses the $Z$ statistic (the maximum absolute difference between the
  cumulative distributions of the samples) to quantify the dissimilarity
  between the distributions of two samples. The larger the value of $Z$,
  the more dissimilar the distributions. \\
  \indent We exclude Haumea and objects with
  $\Delta m$\,$>$\,$0.9$ mag from this calculation:
  Haumea is not representative of the densities of its family, and
  objects with very large $\Delta m$ obey a 
  different relationship between rotational properties and bulk density
  \citep{2007-AJ-133-Lacerda}.
  Considering the two populations made of the 7 family members and
  the 64 background TNOs, we obtain a value of $Z=1.276$.
  The corresponding probability that the $P$ vs. $\Delta m$ distributions
  of family members and other TNOs would differ by more than they do is
  $P_{>Z}=0.040$.
  If we furthermore discard objects with $\Delta m$\,$<$\,$0.1$ mag
  that are unlikely to be Jacobi ellipsoids, the populations are made of
  5 and 42 
  TNOs respectively, and the K-S probability lowers to $P_{>Z}=0.014$.
  These low values of $P_{>Z}$ suggest that the
  family members have different rotational properties from other
  TNOs, although the current data are still insufficient to
  quantitatively compare the densities of family
  members and other TNOs. \\
  \indent We note that the small numbers of objects and rather
  uncertain rotation periods for many, make such an analysis approximate
  at best, \ie, this is not yet a statistically robust result.
  Furthermore, many of the larger objects with long rotation
  periods and low lightcurve amplitudes are likely to be spheroidal
  rather than ellipsoidal bodies, with single peak lightcurves due to
  albedo features (Pluto is an example), and we have made no attempt to
  separate these from the shape controlled bodies in
  Fig.~\ref{fig: dens}. In addition, no restriction on orbit type
  (\eg, classicals, scattered disk) is
  imposed on the objects in Fig.~\ref{fig: dens}, as the total number
  of TNOs with lightcurves in the \citet{2009-AA-505-Duffard}
  compilation is still relatively low (67 objects included in
  Fig.~\ref{fig: dens}).

\section{Family membership and formation scenario\label{sec: discussion}}

 \subsection{Orbital elements} 

\begin{figure}
  \centering
  \includegraphics[width=\columnwidth]{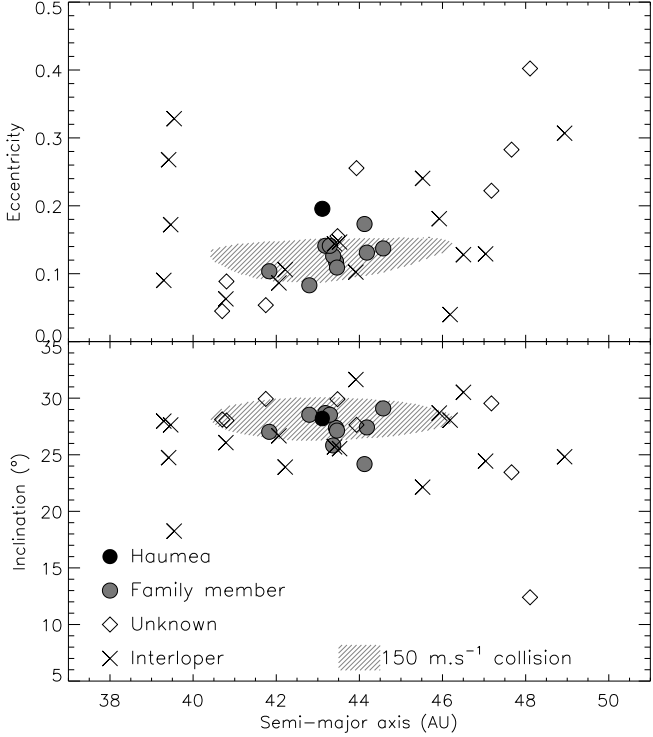} 
  \caption[Orbital parameters of the family candidates]{%
    Confirmed family members
    (grey filled circles with a black outline), 
    rejected candidates: interlopers (crosses),
    and those with unknown surface properties (open diamonds)
    plotted in terms of the orbital osculating parameters
    semi-major axis,
    inclination and eccentricity.
    Haumea itself is shown as a black disk.
    We also drawn the area corresponding to a simulation of
    ejected particules from a nominal collision with
    an isotropic $\Delta v$ of 150\,m\,s$^{-1}$
    \citep{2007-AJ-134-Ragozzine}.
  }
  \label{fig: aei}
\end{figure}

  \indent We show in Fig.~\ref{fig: aei} the 
  orbital parameters (semi-major axis, inclination and
  eccentricity) of the candidates.
  As already noted by \citet{2010-AA-511-Snodgrass}, 
  the confirmed family members cluster tightly around the centre of
  the distribution in both plots, at the supposed location of the
  pre-collision Haumea
  \citep[Haumea itself having now a higher eccentricity, owing to
    its interaction with Neptune through orbital 
  resonance, see][]{2007-AJ-134-Ragozzine}.
  Water ice has been detected on all the objects 
  within the isotropic $\delta v$ limit of 150\,m\,s$^{-1}$ defined
  for a collision-formation scenario by
  \citet{2007-AJ-134-Ragozzine},
  while only 14\% of the objects with a larger velocity dispersion
  harbour water ice surfaces.
  Even assuming that all the as-yet uncharacterised
  candidates have water ice on their surfaces brings this number to
  only 32\%, which significantly differs from the proportion inside
  the 150\,m\,s$^{-1}$ region. 
  The probability of randomly selecting the single most clustered set
  of 11 out of a sample of 36 is only $10^{-9}$.
  The clustering of water-bearing objects around the position of the
  proto-Haumea in orbital parameter space is therefore real, with a
  very high statistical significance. 
  Wider photometric surveys of the trans-Neptunian region 
  \citep{2011-ApJ-730-Trujillo, 2012-ApJ-749-Fraser} find no further
  bodies with the strong water-ice spectrum characteristic of the
  family, which appears to be a unique cluster of objects. \\
%
%
%
%
%
 \subsection{Mass of the family} 

\begin{figure} 
  \centering
  \includegraphics[width=.49\textwidth]{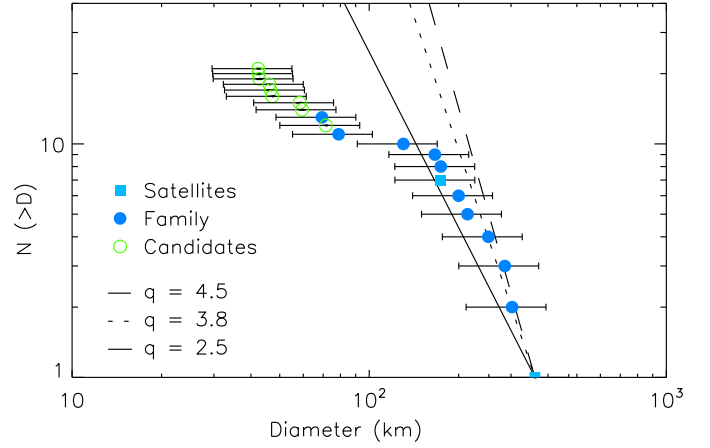} 
   \caption{Cumulative size distribution for confirmed
     (filled blue circles) and
     remaining candidate (open green circles) family members,
     compared with
     three power law models (see text).
     The models have $q=2.5$ (solid line),
     $q=3.8$ (dashed line) and
     $q=4.5$, approximating the model of
     \citet{2010-ApJ-714-Leinhardt}, (dotted line).
     The satellites of Haumea, Hi`iaka and Namaka, are represented
       by blue squares.}
   \label{fig: sfd}
\end{figure}

  \indent We discuss below how current observations can constraint
  the formation scenario of Haumea and its family.
  We first evaluate the mass of the family by summing over all confirmed members.
  We evaluate the mass $M$ of each object from its absolute
  magnitude $H$, from
  \begin{equation}
    M = \frac{\pi \rho}{6} \left( \frac{1329}{\sqrt{p_V}} \right)^3 10^{ -0.6 H},
  \end{equation}

  \noindent where $p_V$ is the geometric albedo (assumed to be 0.7 for
  family members), and $\rho$ their density
  \citep[assumed to be 0.64\,g\,cm$^{-3}$, the largest found for a
    family member, see Fig.~\ref{fig: dens},
    and consistent with the typical density of TNOs,
    see][]{2012-PSS-Carry}.
  The 11 confirmed family members account for only 1\% of the mass of
  Haumea \citep[4\,$\times$\,$10^{21}$\,kg,][]{2009-AJ-137-Ragozzine},
  raising to 1.4\% when also considering Hi`iaka and Namaka, the two
  satellites of Haumea, as family members. 
  Including all the 8 remaining candidates adds only another 0.01\%. \\
  \indent This mass fraction is however a lower limit, as more icy
    family members can be expected to be found. 
    The area encompassed by the
    confirmed family member in orbital element space
    (Fig.~\ref{fig: aei}) is wide (6 AU).
    Given the small fraction of known TNOs
    \citep[a couple of percent, for TNOs of 100\,km diameter, 
      see][]{2008-SSBN-10-Trujillo}, 
    many more objects are still to be discovered in the vicinity of
    Haumea.
    To estimate how much mass has yet to be
    discovered, we compare the observed cumulative size-distribution of
    family members with three simple models, described by power laws of the
    form $N(>r) \propto r^{-q}$ (Fig.~\ref{fig: sfd}).
    The observed
    distribution includes the satellites of Haumea
    (namely Hi`aka and Namaka) which have
    0.29 and 0.14 times Haumea's diameter of 1250\,km
    \citep[][]{2009-ApJ-695-Fraser, 2009-AJ-137-Ragozzine, 2012-PSS-Carry},
    and is based on the observed distribution of absolute magnitudes $H$
    and an assumed Haumea-like albedo of 0.7 (Table~\ref{tab:
      period}), with the exception of 2002 TX$_{300}$,
    which has a diameter determined by stellar
    occultation \citep{2010-Nature-465-Elliot}.
    We also include the remaining
    candidates (open circles) that have not yet been ruled out, which
    are nearly all smaller (fainter) than the confirmed family members. The
    first model is based on the classical distribution for collisional
    fragments, with $q=2.5$ \citep{1969-JGR-74-Dohnanyi}. The second takes the
    size distribution for large TNOs measured by
    \citet{2009-AJ-137-Fraser}, $q=3.8$. The third is a simplification
    of the model presented by \citet{2010-ApJ-714-Leinhardt}, with the mass
    distribution shown in their Fig.~3 approximated by a $q_{\rm M}=1.5$ power
    law, which corresponds to a very steep size distribution of
    $q=4.5$. We normalise the distribution to
    the largest object, Hi`iaka, on the assumption that
    there are no more family members with $H \approx 3$
    ($D \approx 400$\,km) to be found.\\
%
%
%
%
%
    \indent The $q=2.5$ model predicts that the largest object still
    to be discovered has a diameter of around 140\,km, or $H \approx
    5$. This corresponds to an apparent magnitude at opposition fainter
    than 21, which is below the detection limits of wide area TNO
    surveys to date \citep{2003-EMP-92-Trujillo}.
    Extrapolating this model to small sizes predicts a total mass of the family of
    $\sim$2\% of Haumea's mass, with nearly all of that mass in the
    already discovered large fragments.
    Models 2 and 3 predict the largest family members still to be
    discovered of diameters $\sim$220\,km and 250\,km respectively, objects
    at least a magnitude brighter, which would have had a chance of
    being found by existing surveys, depending on where in their orbits
    they currently are.
    These models cannot be extrapolated (model 2 is based on
    the observed TNO size distribution, which has a different slope at
    smaller sizes, and model 3 is a coarse approximation to the
    simulations by \citet{2010-ApJ-714-Leinhardt}, which give a total
    family mass of $\sim$7\% of Haumea), but they do allow there to be
    considerable missing mass in these large undetected bodies. These models
    show that in the case of a collisional size distribution we already know of all 
    the large bodies, and all the significant mass, while steeper distributions can be
    observationally tested as they imply missing members with large
    diameters that should easily 
    be found by new surveys (\eg, Pan-STARRS, LSST). \\
  \subsection{Family formation models}

  \indent The clustering of Haumea's family, with a
    low $\delta v$ between fragments, may be its most peculiar property
    \citep{2011-ApJ-733-Marcus}, and can be used as a strong
    constraint on formation models. 
    Additionally, the models must explain
    the spin of Haumea and the mass and velocity dispersion of its
    fragments, 
    keeping in mind that some of the original mass has been lost over time
    \citep[TNO region is thought to be far less populous today than it
      was in the early solar system, see, \eg,][]{2008-SSBN-5-Morbidelli}.
    None of the models below studied the long-term
    stability of the satellites or the fate of ejected fragment
    formed during the collision/fission, but 
    \citet{2012-MNRAS-421-Lykawka} 
    found that about 25\% of the fragments would not survive over
    4\,Gyr, the first Gyr being when most of the dynamical evolution
    took place. \\
  \indent The model by
  \citet{2009-ApJ-700-Schlichting}, which describes the cataclysmic
  disruption of a large icy satellite around Haumea, 
  reproduces the velocity distribution of the family, and gives an
  original mass of the family of around 1\% of Haumea.
  The spin period of Haumea, however, is expected to 
  be longer than observed, based on considerations on physics of
  impacts and tides in the system
  \citep[see arguments by][and reference
    therein]{2010-ApJ-714-Leinhardt, 2012-MNRAS-419-Ortiz}.
  The rotational fission scenario presented by
  \citet{2012-MNRAS-419-Ortiz} does reproduce Haumea's spin period,
  but predicts a velocity distribution several times higher than
  observed.
  a peculiar kind of \textsl{graze and merge} impact  
  can explain Haumea's shape and spin, and a family of icy objects 
  with low $\delta v$, that have a total original mass $\sim 7\%$ of the 
  proto-Haumea. This mass is higher than that observed, but may be consistent
  with objects lost from the family by dynamical interactions.\\    
  \indent \citet{2011-LPI-42-Cook} suggested an alternative solution,
  that bodies without the unique strong water ice signature could also be
  family members but from different layers in a differentiated proto-Haumea.
  This \textsl{black sheep} hypothesis has fewer observational constraints,
  as currently too few objects are known to be able to identify the family by 
  dynamics alone (\ie, without spectral information), so it is possible to imagine
  a higher mass and larger velocity dispersion.
  However, as discussed above, the clustering of family
  members with icy surfaces suggests that the true family members have
  a small velocity dispersion. Further modelling is required to tell
  whether a low $\delta v$ population of pure ice bodies can come from
  a population of a mixture of higher-velocity collisional fragments. \\

\section{Conclusions}
  \indent We have presented optical and near-infrared colours for 8 of
  the 36 candidate members of Haumea's collisional family
  \citep{2009-AJ-137-Ragozzine}, in addition to the 22 objects we
  already reported \citep{2010-AA-511-Snodgrass}.
  We confirmed the presence of water ice on the surface of
  2003~UZ$_{117}$, confirming its link with Haumea, and rejected 5 other
  candidates
  \citep[following our prediction that most of the
    remaining objects would be interlopers,][]{2010-AA-511-Snodgrass}.\\
  \indent Of the 36 family member candidates including
    Haumea, only 11 (30\%) have been
  confirmed on the basis of their surface properties,
  and a total of 17 have been rejected (47\%). All the
  confirmed members are tightly clustered in orbital elements, the largest
  velocity dispersion remaining 123.3\,m s$^{-1}$ (for 1995~SM$_{55}$).
  These fragments, together with the two satellites of Haumea, Hi`iaka
  and Namaka, account for about 1.5\% of the mass of Haumea. \\
  \indent The current observational constraints on the family
  formation can be summarised as: 
  \begin{enumerate}
  \item A highly clustered group of bodies with unique spectral signatures.
  \item An elongated and fast-rotating largest group member.
  \item A velocity dispersion and total mass lower than expected for a
    catastrophic collision with a parent body of Haumea's size, but a
    size distribution consistent with a collision. 
  \end{enumerate}
  Various models have been proposed to match these unusual
  constraints, although so far none of these match the full set of
  constraints.

\begin{acknowledgements}
  We thank the dedicated staff of ESO's La Silla and Paranal
  observatories for their assistance in obtaining this data.
  Thanks to Blair and Alessandro for sharing their \textsl{jarabe}
  during observations at La Silla.
  This research used VO tools
  SkyBoT \citep{2006-ASPC-351-Berthier}
  and
  Miriade \citep{2008-ACM-Berthier} 
  developed at IMCCE, and 
  NASA's Astrophysics Data System.
  A great thanks to all the developers and
  maintainers. 
  \add{Thanks to an anonymous referee for his comments and careful
    checks of all our tables and numbers.}
  We acknowledge support from the Faculty of the European Space
  Astronomy Centre (ESAC) for granting the visit of C. Snodgrass.
  P. Lacerda is grateful for financial support from a Michael West
  Fellowship and from the Royal Society in the form of a Newton
  Fellowship.
  The research leading to these results has received funding from the
  European Union Seventh Framework Programme (FP7/2007-2013) under grant
  agreement no. 268421.

\end{acknowledgements}

\bibliographystyle{aa}      
\bibliography{biblio}

\end{document}